\documentclass[prb, superscriptaddress,reprint,amsmath,longbibliography]{revtex4-2}
\usepackage{graphicx}
\usepackage{amssymb}
\usepackage{physics}
\usepackage{hyperref}
\usepackage{refcount}
\usepackage{comment}
\usepackage{caption}
\usepackage{ragged2e}
\usepackage[utf8]{inputenc}

\hypersetup{pdfstartview=FitH,pdfpagemode=UseOutlines,colorlinks,
citecolor=blue,linkcolor=blue,urlcolor=blue}

\usepackage[format=plain,justification=raggedright]{caption}

\begin{document}
\title{Interband State Transfer in Double-Gated Bilayer Graphene at High Electric Field}

\author{Margherita Melegari}
    \email[]{margherita.melegari@unige.ch}
	\affiliation{Department of Quantum Matter Physics, University of Geneva, Quai Ernest-Ansermet 24, 1211 Geneva, Switzerland}
    \affiliation{ Department of Applied Physics, University of Geneva, 24 Quai Ernest Ansermet, Geneva, CH-1211 Switzerland}
    \author{Brian Skinner}
    \affiliation{Department of Physics, Ohio State University, Columbus, OH, USA.}
    \author{Ignacio Gutierrez Lezama}
    \email[]{ignacio.gutierrez@unige.ch}
    	\affiliation{Department of Quantum Matter Physics, University of Geneva, Quai Ernest-Ansermet 24, 1211 Geneva, Switzerland}
    \affiliation{ Department of Applied Physics, University of Geneva, 24 Quai Ernest Ansermet, Geneva, CH-1211 Switzerland}

    \author{Alberto F. Morpurgo}
    \email[]{alberto.morpurgo@unige.ch}
 	\affiliation{Department of Quantum Matter Physics, University of Geneva, Quai Ernest-Ansermet 24, 1211 Geneva, Switzerland}
    \affiliation{ Department of Applied Physics, University of Geneva, 24 Quai Ernest Ansermet, Geneva, CH-1211 Switzerland}

\date{\today}

\begin{abstract}
The band structure of Bernal-stacked bilayer graphene can be tuned using double-gated transistors to apply a perpendicular electric field that generates an interlayer potential energy difference $\Delta$. Dielectric breakdown limits the operation of conventional devices to the $\Delta \ll t_\perp \simeq 360$ meV regime. We employ double ionic gating to reach  fields past  $ 1$ V/nm, for which $\Delta > t_\perp$. We find that for $\Delta \simeq t_\perp$, the evolution of the longitudinal resistance ($R_{xx}$) peak as a function of applied gate voltages undergoes a sharp change in slope, exhibiting a pronounced ``knee". Increasing $\Delta$ past the ``knee" results in an unusual evolution  transport properties: the peak in $R_{xx}$ decreases in magnitude, it  exhibits a splitting concomitant with multiple  sign reversals of the Hall resistance, and  hysteresis in the peak position emerges. We explain the observed phenomenology in terms of in-gap bound states, whose energy strongly depends on the perpendicular  electric field, and crosses the mid-gap level for sufficiently large $\Delta > t_\perp$. The phenomenon causes large changes in the electronic density of in-gap states that  profoundly affect the evolution of the chemical potential. Our experimental results and their interpretation reveal unique aspects of the physics of in-gap states in Bernal bilayer graphene and demonstrate that double ionic gating enables investigating the large-$\Delta$ regime, which has remained experimentally inaccessible so far.
\end{abstract}

\maketitle

\section{Introduction}
The band structure of Bernal-stacked bilayer graphene (BLG) can be controlled  by applying a perpendicular electric field to establish an interlayer electrostatic potential difference, which generates a difference in energy $\Delta$ for electrons residing in the two layers \cite{mccann_landau-level_2006,mccann_asymmetry_2006,castro_biased_2007,min_ab_2007,oostinga_gate-induced_2008,zhang_determination_2008,zhang_direct_2009,kuzmenko_determination_2009,mak_observation_2009,weitz_broken-symmetry_2010, young_electronic_2012,velasco_transport_2012,mccann_electronic_2013}. If $\Delta=0$, BLG is a zero-gap semiconductor with valence and conduction bands that touch at isolated points in the Brillouin zone. A finite perpendicular electric field opens up a gap equal to $\Delta$, giving electrostatic control over the band structure. The possibility to modify the band structure in this way has a multitude of implications, including the realization of gate-defined nanostructures \cite{allen_gate-defined_2012,goossens_gate-defined_2012,varlet_anomalous_2014,overweg_electrostatically_2018,overweg_topologically_2018,eich_spin-valley_2018,kurzmann_charge_2019,duprez_spin-valley_2024}, the ability to induce Berry curvature and control the band topology near the $K/K'$ points \cite{xiao_valley-contrasting_2007,martin_topological_2008,shimazaki_generation_2015,yin_tunable_2022}, and much more. 

\begin{figure}[h!]
    \centering
    \includegraphics[width=0.9\columnwidth]{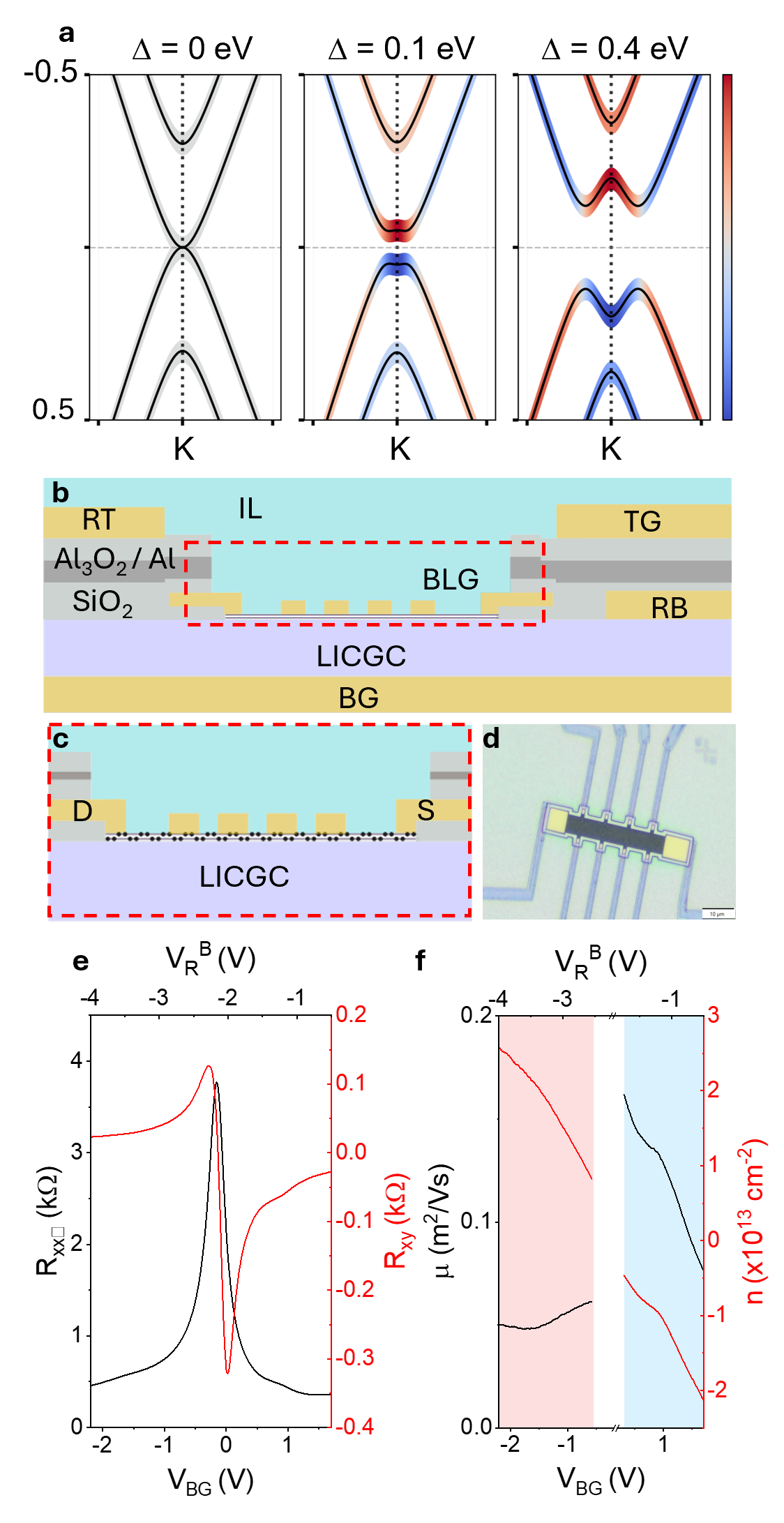}
    \caption{\justifying\small (a) Band structure of BLG for different values of interlayer energy difference $\Delta  = 0$, $0.1$, and $0.4$ eV (the color scale shows how the layer-resolved weight of the electronic states varies on the different bands, as a function of energy). (b) Schematic cross-section (not to scale) of a BLG transistor equipped with an IL top gate and a Li-ion glass-ceramic back gate, together with their corresponding reference electrodes (RT, RB). The SiO$_2$/Al/Al$_2$O$_3$ trilayer is used to electrostatically decouple the top and bottom electrolytes away from the BLG. (c) Zoom-in on the device channel area, showing the Pt electrodes contacting the BLG. (d) Optical image of a  device patterned in a Hall bar geometry on a LICGC substrate, prior to the deposition of the ionic liquid (IL). (e) Longitudinal ($R_{xx\square}$, black curve) and transverse ($R_{xy}$, red curve) resistance measured as a function of gate voltage on a Li-ion–gated BLG device, prior to the deposition of the ionic liquid ($R_{xy}$ is measured by applying a perpendicular magnetic field $B=1$ T). The top axis shows the value of the reference potential measured while sweeping the voltage applied to the backgate. (f) Mobility $\mu$ and carrier density $n$ extracted from the measurements shown in panel (e). }
    \label{fig:Figure1}
\end{figure}

In experiments, the  interlayer potential difference is normally applied using double-gate devices \cite{mccann_landau-level_2006,mccann_asymmetry_2006,castro_biased_2007,min_ab_2007,oostinga_gate-induced_2008,zhang_determination_2008,zhang_direct_2009,kuzmenko_determination_2009,mak_observation_2009,weitz_broken-symmetry_2010, young_electronic_2012,velasco_transport_2012,mccann_electronic_2013}, in which two  gate  electrodes on opposite sides of a BLG give independent control over the perpendicular electric field and the chemical potential (i.e, the accumulated charge density). The maximum size of the band gap that can be opened in this way is limited by the electric breakdown of the insulators that separate the BLG from the gate electrodes, and recent experiments --together with the re-analysis of old ones-- have shown that the largest values reached in actual devices are less than 200 meV \cite{slizovskiy_out-of-plane_2021,tenasini_band_2022}. This limitation has so far prevented exploring experimentally the regime $\Delta > t_\perp$ ($t_\perp \simeq 360$ meV is the interlayer hopping integral), in which theory predicts a complete reshaping of the band structure (see Fig \ref{fig:Figure1}a). For $\Delta > t_\perp$, the bandgap should stop increasing linearly and saturates at  $t_\perp$; the band  edges are expected to move away from the $K/K'$ points \cite{mccann_landau-level_2006, koshino_electron_2008,mccann_electronic_2013}, causing an unusual Berry curvature distribution in $k$-space; the dispersion relation becomes ``inverted", leading  to a so-called ``moat" band accompanied by a divergent density of states that can strongly enhance interaction effects \cite{koshino_electron_2008,skinner_interlayer_2016,skinner_bound_2014,mccann_landau-level_2006}. For $\Delta \gg t_{\perp}$ the energy accompanying this band reshaping is large --hundreds of meV, much larger than that associated to the so-called Mexican-hat dispersion present at very small $\Delta$-- so that the accompanying physical phenomena are expected to be robust. However, the limited perpendicular electric field that could be applied so far has prevented  exploring the occurrence of new electronic phenomena that may occur in the large interlayer potential regime.

Here we report transport measurements on BLG devices equipped with double ionic gates, which allow the application of perpendicular electric fields as large as 3 V/nm \cite{domaretskiy_quenching_2022}, sufficient to reach deep into the  $\Delta>t_\perp$ regime \cite{mccann_landau-level_2006, koshino_electron_2008,mccann_electronic_2013}. We first show that --when operated in the low interlayer potential regime-- our double ionic gated devices exhibit the established behavior, with the peak in longitudinal resistance that shifts linearly when the two gate electrodes are biased with opposite polarity, while simultaneously increasing in magnitude due to the opening of a bandgap \cite{castro_biased_2007,oostinga_gate-induced_2008,zhang_determination_2008,mak_observation_2009, zhang_direct_2009,kuzmenko_determination_2009}. When the estimated electric field in the BLG is increased past approximately 1 V/nm --corresponding to $\Delta \simeq 400$ meV-- the observed behavior changes drastically. The evolution of the resistance peak with both gate voltages starts deviating pronouncedly from linearity, the maximum resistance decreases upon increasing the interlayer potential, the resistance peak exhibits a splitting, and hysteresis in its position appears. We propose a scenario that explains this phenomenology in terms of  in-gap states of electrons (holes) bound to positive (negative) ions in the top electrolyte that-- when  $\Delta$ is increased past $t_\perp$-- shift rapidly in energy and cross the mid-gap level (i.e., they are transferred from one band to the other). The resulting modification of the in-gap density of states strongly affects the  evolution  of the chemical potential in a way that accounts for the experimental observations. Our results show that the double ionic gated devices allow generating an interlayer potential significantly larger than $t_\perp$ and enable the investigation of the electronic properties  of this yet unexplored regime.

\begin{figure*}[ht]
    \centering
    \includegraphics[width=\textwidth]{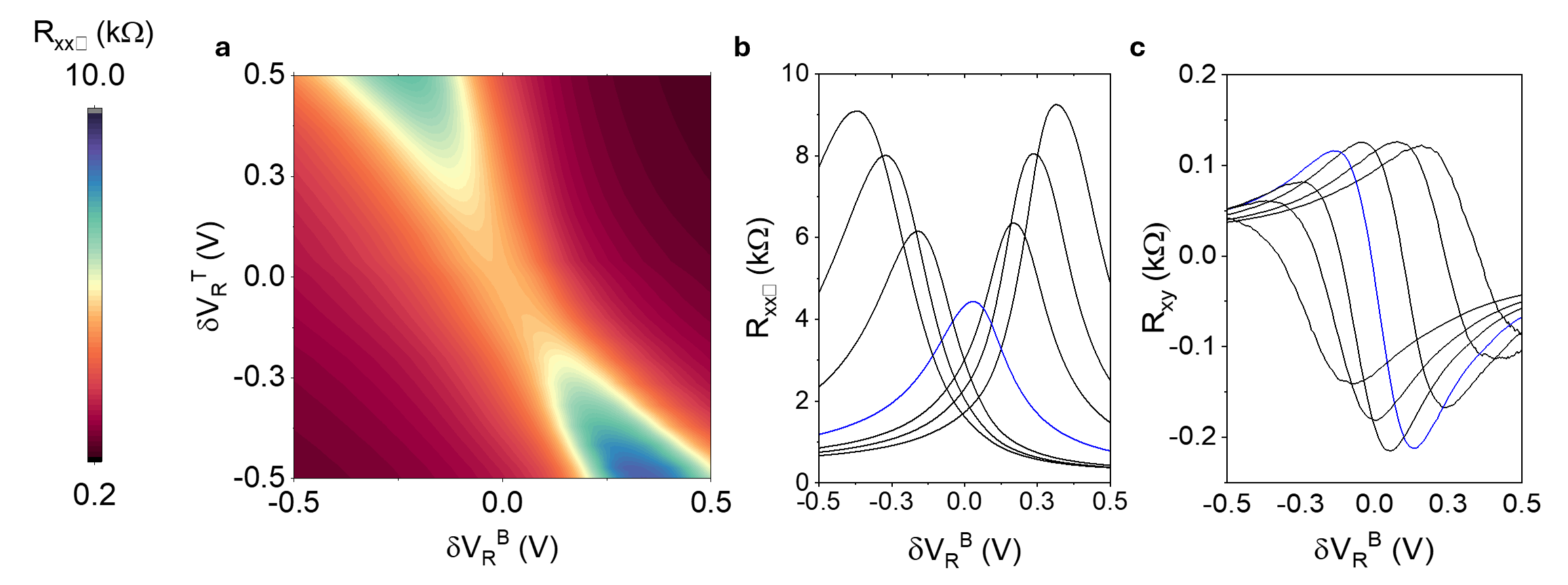}
    \caption{\justifying\small\textbf(a) Color plot of the longitudinal square resistance $R_{xx\square}$ as a function of top and bottom reference potentials ($\delta V_R$) relative to the reference potential measured at zero perpendicular electric field (see main text). In this range of applied electric field, the data exhibit the characteristic evolution expected  for a double-gated BLG regime: the maximum of $R_{xx}$ shifts linearly when $\delta V_R^T$ and $\delta V_R^B$ are varied with opposite polarity, and the peak resistance increases with perpendicular electric field (proportional to $\delta V_R^T - \delta V_R^B$). (b) Selected cuts of the color map in panel (a) for different fixed values of $\delta V_R^T$ (the blue curve of minimum peak resistance corresponds to the trace used to define the zero–electric-field point, $V_{R_0}^T$ and $V_{R_0}^B$ at the position of the peak). (c) Corresponding evolution of the transverse resistance $R_{xy}$ measured in the presence of a $B=1$ T perpendicular magnetic field.}
    \label{fig:Figure2}
\end{figure*}

\section{Transport measurements on double ionic gated bilayer graphene devices}
Fig.~\ref{fig:Figure1}b, \ref{fig:Figure1}c and \ref{fig:Figure1}d show the structure of double ionic gated transistors and an optical microscope image of the core part of an actual device. A BLG etched into a Hall-bar is in direct contact with two independent electrolytes coupled to its top and bottom layers. The top electrolyte consists of an ionic liquid (we employ DEME-TFSI) \cite{fujimoto_electric-double-layer_2013} and the back electrolyte is a Li-ion glass ceramic substrate (LICGC, AG01 purchased from O'Hara Corporation). The two electrolytes are separated by a grounded metallic layer (consisting of  SiO$_{2}$/Al/Al$_{2}$O$_{3}$) that prevents their direct electrostatic coupling away from the BLG. Two large evaporated metal pads --each in contact with one of the two electrolytes-- are used to apply the corresponding gate voltage. Because of the extremely short electrostatic screening length, the applied gate voltage drops at the interfaces between gate and electrolyte and between electrolyte and the device (i.e., there is no voltage drop inside the electrolyte)\cite{shimotani_electrolyte-gated_2006,CarrierLoc_Frisbee,fujimoto_electric-double-layer_2013,bisri_endeavor_2017}, and the voltage across each electrolyte/device interface is monitored using so-called reference electrodes (see electrodes labeled ``RT" and ``RB" in the Fig.~\ref{fig:Figure1}b) \cite{gutierrez-lezama_ionic_2021}. All transport measurements discussed here were performed at (or close to) room temperature, for which the ions in the electrolytes are mobile. Several details of the device configuration and of the measurements protocols are essential for a correct, stable and reproducible device operation. We refer the interested reader to our earlier work for a detailed discussion of these  technical aspects~\cite{gutierrez-lezama_ionic_2021, cao_full_2023, domaretskiy_quenching_2022}.

It is important to appreciate that the two electrolytes employed in our devices are different. In the top ionic liquid, both the positive (DEME$^+$) and negative (TFSI$^-$) ions are free to move and respond to changes in the gate voltage. Both ions are therefore present at the ionic liquid/BLG interface irrespective of the magnitude and polarity of the gate voltage, which only determines their relative concentration \cite{bisri_endeavor_2017,fujimoto_electric-double-layer_2013}. In the Li-ion electrolyte substrate, on the other hand, only the positive Li-ions are mobile, as the compensating negative charge is fixed in the glass matrix \cite{lin_progress_2023}. As a result, for large gate voltage applied to the back gate only charge of one polarity is present next to the BLG: positive Li-ions for large positive back gate voltages and the negative charge in the glass matrix for large negative gate voltages. As we will discuss later, this asymmetry is important to understand the behavior of the devices at large applied perpendicular electric field. Note also that for large negative gate voltage applied to the back gate we should expect that large potential fluctuations are imprinted in the BLG, because the negative charge in the glass matrix cannot redistribute spatially to lower its electrostatic energy by creating a more uniform potential. 

As a first step, we characterize our BLG devices as a function of back gate voltage $V_{BG}$ by measuring them prior to depositing the ionic liquid, because not much is known about the operation of devices with Li-ion glass substrates (in contrast to ionic liquid electrolytes, which have been employed extensively in the past) \cite{ye_accessing_2011, efetov_controlling_2010}. Fig.\ref{fig:Figure1}e shows the square longitudinal resistance $R_{xx\square}$ (black curve) and the Hall resistance $R_{xy}$ (red curve) measured at room temperature, from which we extract the accumulated carrier density and mobility (red and black curves in  Fig.~\ref{fig:Figure1}f, respectively). The application of approximately $\pm 2$ V results in the accumulation of 2-3 10$^{13}$ carriers per cm$^2$ \cite{ye_accessing_2011}, a value comparable to that reached with ionic liquid electrolytes for the same applied gate voltage \cite{ye_accessing_2011} (whenever possible, with Li-ion glass we refrain from applying gate voltages in excess of $\pm$ 2.5 V, which increases the possibility of device failure). The carrier mobility ($\mu \approx 10^3$ cm$^2$/Vs) is also comparable to that measured with ionic liquid electrolyte gates \cite{ye_accessing_2011}(the hole mobility is somewhat lower than the electron mobility, likely due to the larger potential fluctuations present for negative applied back gate voltage, as just mentioned above).

\begin{figure*}[ht]
    \centering
    \includegraphics[width=\textwidth]{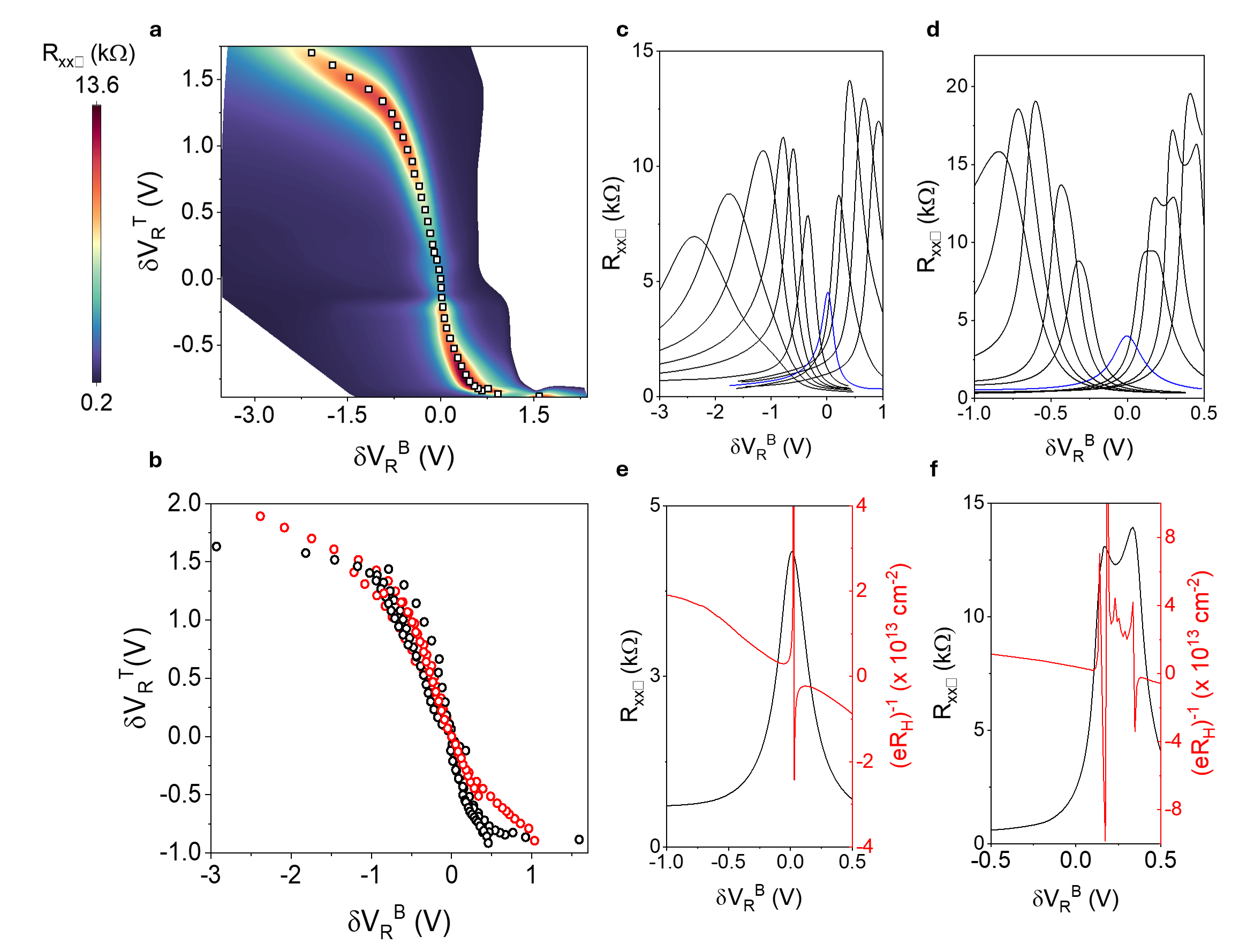}
    \caption{\justifying\small Color plot of $R_{xx\square}$ as a function of $\delta V_R^B$ and $\delta V_R^T$ extending to the high electric-field regime. The evolution of the peak position (marked by the white squares) strongly deviates from linearity: after an initial linear trend, it exhibits  a clear ``knee", and eventually flattens at large $\delta V_R^T$. (b) The evolution shown in panel (a) is reproducible across multiple devices and using different pairs of contacts. The panel contains seven distinct data sets measured using multiple pairs of contacts on two different devices (symbols  of different colors represent data taken on the two different devices).  (c) Selected line cuts extracted from panel (a), showing $R_{xx\square}$ as a function of $\delta V_R^B$ for fixed different values of $\delta V_R^T$. The height of the peak in $R_{xx\square}$ decreases once $\delta V_R^T$ is increased past the ``knee", for both polarities of perpendicular electric field. (d) measurements of $R_{xx\square}$ as in panel (b) performed using a different pair of voltage probes as compared to those used to take the data in (a). Irrespective of the probes,  the peak in $R_{xx\square}$  broadens substantially for negative values of $\delta V_R^B$ past the ``knee", whereas for positive $\delta V_R^T$ the peak remains narrower, and --when using probes for which the peak is narrowest-- a splitting is detected. (e,f) $R_{xx\square}$ and Hall resistance $R_{xy}$ as a function of $\delta V_R^B$ in the low-and high–electric-field regimes, respectively ($R_{xy}$ is measured in the presence of a $B=1$ T perpendicular magnetic field). In the low-field regime the Hall effect behaves as expected, with the zero crossing occurring at the position of  the maximum in $R_{xx\square}$. In the high-field regime, when a splitting in the peak of $R_{xx\square}$ is seen, $R_{xy}$ changes sign multiple times.  }
    \label{fig:Figure3}
\end{figure*}

The top $x$-axis of Figs.~\ref{fig:Figure1}e and \ref{fig:Figure1}f reports the value of the reference potential $V_R^B$, measured while sweeping the gate voltage $V_{BG}$. Plotting data versus reference potential -- which measures the actual voltage drop at the electrolyte/BLG interface-- allows us to eliminate spurious effects that take place at the gate/electrolyte interface and to limit bias stress  \cite{gutierrez-lezama_ionic_2021}. Specifically, in the following we plot our data as a function of $\delta V_R$, the difference between the reference potential and the reference potential measured at the $R_{xx\square}$ maximum, in the absence of perpendicular electric field (blue traces in Fig.\ref{fig:Figure2}(b) and in Figs \ref{fig:Figure3}(c) and \ref{fig:Figure3}(d)). Following this procedure eliminates sample-to-sample variations in built-in field, drastically minimizes bias stress effect, and enables us to compare quantitatively data measured on different devices over long periods of time (which is essential as measurements on a device can easily take as long as 1 to 2 months).

Fig.~\ref{fig:Figure2}a shows the square resistance $R_{xx\square}$ as a function of $\delta V_R^T$ and $\delta V_R^B$ in the regime where the perpendicular electric field is  smaller than $\Delta$. Probing this well-established regime is important to check if the devices behave as expected. This is indeed the case. The position of the maximum in $R_{xx\square}$ evolves linearly upon varying $\delta V_R^T$ and $\delta V_R^B$ with opposite polarity (see the color plot in Fig.~\ref{fig:Figure2}a). The peak resistance increases as the perpendicular electric field (proportional to $\delta V_R^T-\delta V_R^B$) increases, consistently with the  opening of a band gap between conduction and valence band. Selected cuts of the color map  are plotted in Fig.\ref{fig:Figure2}b for fixed values of $\delta V_R^T$. Fig.~\ref{fig:Figure2}c shows the corresponding evolution of $R_{xy}$. In a few cases we have measured the temperature dependence of the peak $R_{xx\square}(\delta V_R^B)$ value in the presence of perpendicular electric field, and found a weakly insulating behavior. This is consistent with the opening of a gap and the presence of a large density of in-gap states induced by the disorder caused by the ions in the electrolytes that are in direct contact with the BLG (as expected). Measurements in the small electric field regime therefore confirm that double ionic gated devices operate properly, and can be used to investigate the so far unexplored large electric field regime.

\begin{figure}[t]
    \centering
    \includegraphics[width= \columnwidth]{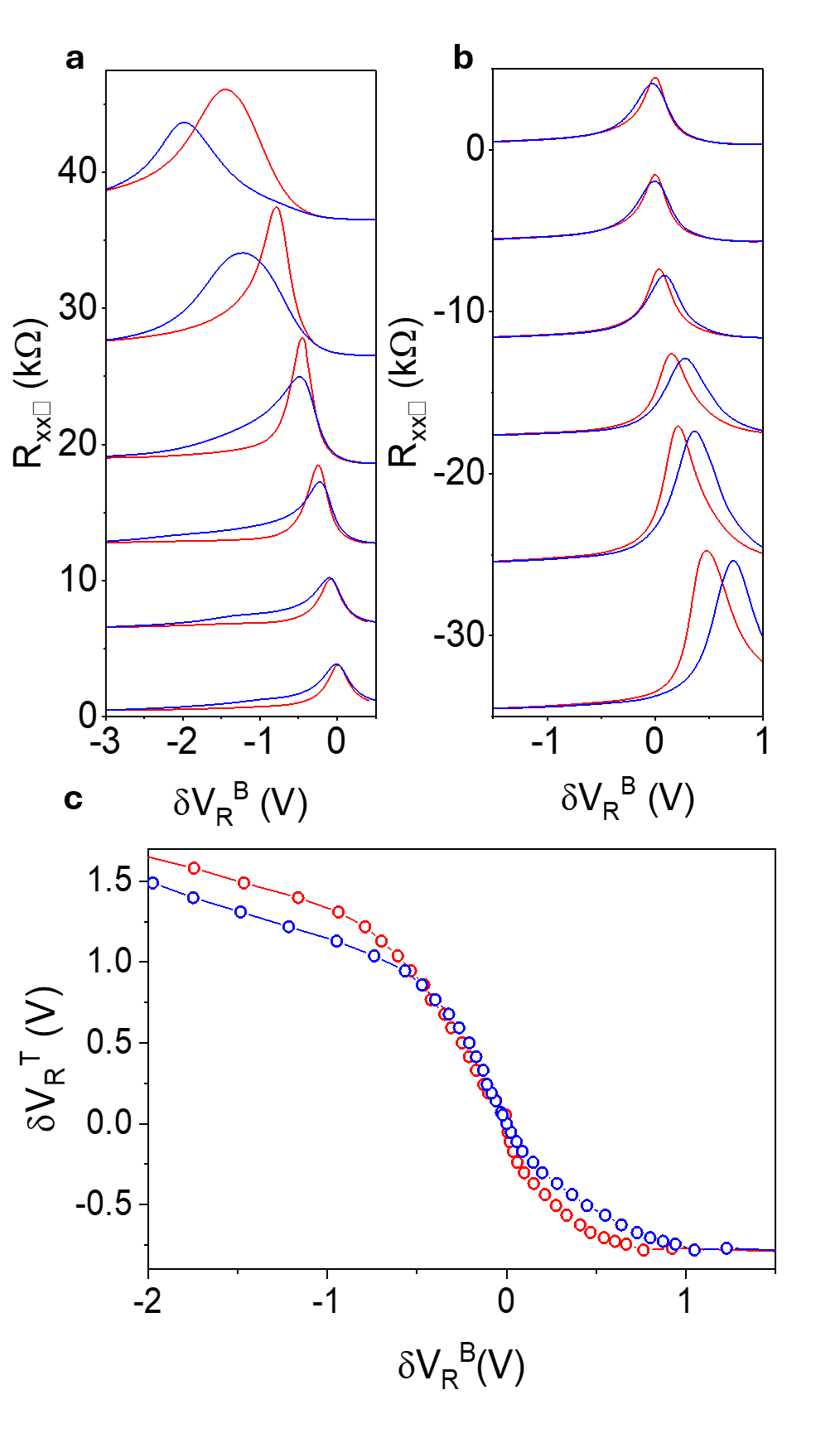}
    \caption{\justifying\small (a,b) $R_{xx\square}$ measured for different values of $\delta V_R^T$ [(a) positive and (b) negative; in both panels the curves are offset vertically for clarity],  while sweeping the back-gate voltage up (from negative to positive values, red curve) and down (from positive to negative values, blue curve). For small $\delta V_R^T$ (i.e., when $\delta V_R^T$ is below the “knee”), the peak position of $R_{xx\square}$ measured in the up- and down-sweeps coincides. For large magnitudes of $\delta V_R^T$ (above the “knee”), a sizable hysteresis in position of the  $R_{xx\square}$ peak is observed.  (c) Position of the peak in $R_{xx\square}$ extracted from up- (red symbols) and down- (blue symbols) sweeps, as a function of $\delta V_R^T$ and $\delta V_R^B$. The onset of hysteresis in the peak position coincides with the ``knee" in the evolution of $R_{xx\square}$.}
    \label{fig:Figure4}
\end{figure}

The evolution of the longitudinal and transverse resistance measured in a much larger range of applied gate voltages exhibits a drastically different phenomenology, whose key aspects are illustrated in Fig.~\ref{fig:Figure3} and Fig.~\ref{fig:Figure4}. The dependence of the  $R_{xx\square}$ peak position on $\delta V_R^T$ and $\delta V_R^B$ strongly deviates from linearity (see Fig.~\ref{fig:Figure3}a): for both polarities, the linear evolution observed at small electric field exhibits a ``knee" and eventually flattens. The behavior is fully reproducible as shown by 7 different data sets,  measured on two different devices, using different pairs of probes (see Fig.~\ref{fig:Figure3}b; red and black symbols corresponds to data taken on two different devices). Note how the ``knee" occurs at virtually identical values of  $\delta V_R^T$ and $\delta V_R^B$ in the two different devices studied, which indicates the deterministic nature of the phenomenon. The height of the peak in $R_{xx\square}(\delta V_R^B)$ measured for different values of $\delta V_R^T$ initially increases, and eventually decreases as  $\delta V_R^T$ is biased past the ``knee" (Fig.~\ref{fig:Figure3}c). For negative $\delta V_R^B$, the peak becomes broad (again,  due to the negative charges in the Li-ion glass that cause larger potential fluctuations). For positive $\delta V_R^B$ the peak remains sharp, which enables a splitting to be detected when  $\delta V_R^T$ is increased past the ``knee" (Fig.~\ref{fig:Figure3}d). Note how, at low perpendicular electric field (when  the peak is not split), the Hall resistance exhibits the usual behavior, changing sign once in correspondence of the maximum of the  $R_{xx\square}(\delta V_R^B)$ curve (Fig.~\ref{fig:Figure3}e). Instead, at larger electric field --when the peak splits-- the Hall effect exhibits a more complex behavior in the $R_{xx\square}$ peak region, changing sign three times, as if  the chemical potential would cross the charge neutrality level multiple times as $\delta V_R^B$ is swept across the peak (Fig.~\ref{fig:Figure3}f). 

One last phenomenon that occurs reproducibly when $\delta V_R^T$ is increased past the ``knee" is hysteresis in the position of the peak of $R_{xx\square}(\delta V_R^B)$ (see Fig.\ref{fig:Figure4}) when sweeping the back gate voltage up and down (i.e., from negative to positive voltages and back). For both polarities of $\delta V_R^T$, the position of the peak in $R_{xx\square}(\delta V_R^B)$ coincides when sweeping up (red curves) and down (blue curves)  $\delta V_R^B$, as long as the device is biased below the ``knee" (i.e., if $\delta V_R^T$ is sufficiently small). If the electric field is large (i.e., if $\delta V_R^T$ is fixed past the ``knee") a clear difference in peak position is visible when sweeping $\delta V_R^B$ up and down. The position of the peak in $R_{xx\square}$ as a function of  $\delta V_R^T$ for up (red symbols) and down (blue symbols) sweeps of  $\delta V_R^B$ is summarized in Fig.~\ref{fig:Figure4}c, where it is apparent that the onset of the hysteresis coincides with the ``knee" in the evolution of $R_{xx\square}$ with  $\delta V_R^T$ and $\delta V_R^B$. All the  unusual aspects of the response of BLG shown in Fig.~\ref{fig:Figure3} and Fig.~\ref{fig:Figure4} are different manifestations of a same microscopic regime, as they all appear for the same intervals of $\delta V_R^T$ and $\delta V_R^B$, when the perpendicular electric field is large.

\section{The role of in-gap bound states}
To understand the origin of the observed experimental observations we start by estimating the magnitude of the electric field $E_{BLG}$ --the electric field present between the two graphene layers forming the BLG-- and  the corresponding interlayer potential $\Delta= E_{BLG} d$ ($d=0.34$ nm is the thickness of bilayer graphene) at the ``knee", i.e. at the point  when the evolution of the position of the peak resistance as a function of $\delta V_R^T$ and $\delta V_R^B$ deviates from linearity. $E_{BLG}$ is given by the difference between the applied perpendicular electric field $E_{ext}$, and the screening  field generated by the density of charge $\delta n$ displaced from one layer to the other:
\begin{equation}
    E_{BLG} = E_{{ext}} - \frac{e\,\delta n}{\varepsilon\,\varepsilon_0},
    \label{eq:EBLG}
\end{equation}
where $\varepsilon\approx 3$ is the relative dielectric constant for BLG \cite{slizovskiy_out-of-plane_2021}. We consider that the device biased just before the ``knee" is still  in  the small field regime, and assume that in this regime  the established theory of the electronic properties of BLG can be employed to estimate  $\delta n = \delta n(E_{BLG}) \simeq \alpha E_{BLG}$, with $\alpha \approx 1.34\times10^{8}\ \mathrm{V^{-1}\,m^{-1}}$ (the result can be obtained by a simple tight-binding calculation)\cite{mccann_asymmetry_2006,mccann_electronic_2013}. To calculate $E_{ext}$, we consider that the applied potential difference  $\delta V_R^T-\delta V_R^B$ drops across a total distance given by the thickness of the BLG plus the thickness of the screening layer of the two electrolytes. From the known geometrical capacitance ($\approx 50 \mu$F/cm$^2$) we estimate the thickness of these screening layers to be 0.1 nm (i.e., the ions are virtually in direct contact with the BLG). From Eq. \ref{eq:EBLG} we then find that --at the  ``knee"-- $E_{BLG} \approx 1.2-1.3\ \mathrm{V/nm}$, corresponding to an interlayer energy difference  of $\Delta \approx 0.42\ \mathrm{eV}$ (slightly larger than $t_\perp$). The unusual phenomenology that we observe is therefore indeed the manifestation of the large electric field regime.

Finding that at the ``knee" $\Delta$ is comparable to $ t_\perp$ accounts for the observed evolution of the peak value of $R_{xx\square}$, namely for the observation that past the ``knee" the peak of $R_{xx\square}$ decreases with further increasing the perpendicular electric field. The reason is that in a double ionic gated device, applying compensating gate voltages of opposite polarities has two effects. The first is to open  a band gap while keeping the chemical potential at charge neutrality, which suppresses the conductance. The second is to introduce disorder in graphene as a result of the proximity of the ions that generate in-gap states in the BLG, which enhance the conductance at charge neutrality. In the small electric field regime, increasing the gate voltages leads to a larger electric field and hence a larger gap. This effect  dominates over the increase in the density of in-gap states, which is why the peak resistance increases. When the interlayer potential $\Delta$ becomes larger than $t_\perp$, however,  the gap saturates and further increasing the gate voltages only adds more ions next to the BLG (and hence more states in the gap). This is why in this regime  the height of the peak in $R_{xx\square}$ decreases. 

In-gap bound states can also qualitatively account for the rest of the observed phenomenology. As mentioned earlier, in the top ionic liquid gate, ions of both polarities are present next to the BLG. When electrons or holes are accumulated on the top layer, they  can bind to ions of opposite polarity and form in-gap  states. The evolution of the energy of these bound states with interlayer potential has been investigated theoretically in the past\cite{skinner_bound_2014, gorbar_electron_2024}. While the precise energy of this bound state depends on the dielectric constant of the medium around the BLG and on the distance of the ionic charge that creates the bound state, its qualitative evolution with $\Delta$ does not. In particular, for $\Delta > t_\perp$, the energy of the lowest in-gap bound state evolves with increasing $\Delta$, and dives deeper inside the band gap (away from the band edge from which bound state originates). Eventually, for sufficiently large $\Delta$, the bound state crosses the mid-gap energy, so that in-gap bound states that are normally empty (if the chemical potential $\mu=0$ is in the center of the gap) become filled with electrons\cite{skinner_bound_2014}. Such a shift across the mid-gap level corresponds effectively to a transfer of states from the conduction to the valence band and is not considered by the conventional description of the electronic properties of BLG, which  neglects the presence of ionic impurities and in-gap states. In double ionic gated devices, however, such in-gap states have a large influence on the evolution of $\mu$.  

To understand how the evolution of the in-gap state energy with interlayer potential explains our experimental observations, let us first discuss the physics qualitatively. Let us imagine to bias the device  just before the ``knee" by applying a positive  $\delta V_R^B$ and a negative $\delta V_R^T$, with the chemical potential near charge neutrality (i.e., with the device biased at the $R_{xx\square}$ peak). An additional increase in $\delta V_R^B$ has two effects. It adds electrons, thereby pushing the chemical potential away from charge neutrality towards the conduction band (since we consider that there are states inside the gap). It also increases the interlayer potential, thereby causing the in-gap bound states to shift to lower energy, with the accompanying change in density of states that pushes the chemical toward lower energies. This last effect occurs at large perpendicular electric fields and is strongly non-linear. The ``knee" corresponds to  the condition for which the in-gap bound state energy crosses the middle of the band gap, so that at $\mu = 0$ empty in-gap states become filled. As a result,  the chemical potential increases much less than expected from conventional theory upon changing  $\delta V_R^B$, which directly implies  that the potential $\delta V_R^T$ on the other gate electrode also has to be changed much less to maintain the chemical potential in the middle of the gap. This is precisely what is  observed in the experiments, namely that the evolution of the $R_{xx\square}$ peak position as a function of  $\delta V_R^T$ and $\delta V_R^B$ flattens. 

When the density of in-gap states is sufficiently large --which is the case in ionic gated devices where the ions are densely packed next to the BLG-- the chemical potential $\mu$ can even decrease when the gate voltage is increased to add electrons.  The splitting in the peak of $R_{xx\square}(\delta V_R^B)$ and the concomitant multiple changes in sign of the Hall resistance (see Fig.~\ref{fig:Figure3}c and Fig.~\ref{fig:Figure3}e) provide direct evidence that in our experiments the  chemical potential does shift from above to below the middle of the gap when $\delta V_R^B$ is increased past the ``knee" and more electrons are  added to the system. A negative change in $\mu$  upon adding electrons means that $d\mu/dn <0$, and indicates the tendency of the system to become unstable. To be precise, for an isolated system thermodynamic  stability at equilibrium requires $d\mu/dn>0$, but in the presence of countercharges accumulated by a gate electrode, the stability condition is modified into $1/C_g +d\mu/dn >0$ \cite{skinner_anomalously_2010}, where $C_g$ is the gate capacitance. Yet, in practice, finding that $d\mu/dn <0$ indicates at very least the presence of strong positional correlations between electrons \cite{skinner_anomalously_2010}, so that increasing the electron density is associated with large-scale rearrangement of electron positions, suggesting a tendency towards instability. (With the very large geometrical capacitance of ionic gates, $1/C_g$ is a small quantity and it may easily be that in our devices $1/C_g +d\mu/dn <0$).

These considerations are relevant for our experiments, because  upon sweeping the gate voltage a large density of the electrons has to rearrange to re-equilibrate. In the presence of disorder responsible for the in-gap states, such a re-equilibration is likely a slow process, and may not occur completely on the time scale of our measurements when changing the sweep direction of $\delta V_R^B$. The  hysteresis in the position of the  peak of $R_{xx\square}(\delta V_R^B)$ that we observe (Fig.\ref{fig:Figure4}) is a direct consequence of this fact (i.e., on the time scale of the measurements, at large perpendicular electric field the state of the system is not the same upon sweeping $\delta V_R^B$ up or down). The presence of hysteresis  is therefore fully consistent with the scenario that we propose to explain the unconventional phenomenology observed in BLG at large interlayer potential.

\section{Modeling the evolution of  the chemical potential}
\label{sec:modeling}
To better appreciate the qualitative considerations just made, we describe the evolution of the chemical potential as a function of top and back gate voltages using a simple model of in-gap bound states. Our goal is not to precisely reproduce  the experimental results -- which depend on too many unknown microscopic details -- but only to illustrate by means of calculations performed under well-defined conditions that the key aspects of the evolution of the chemical potential in the presence of in-gap bound states indeed conform to the qualitative considerations made above and reproduce the correct orders of magnitude of the measured quantities. 

To this end, we first consider the in-gap bound state created by a single, isolated, coplanar impurity charge. When a single charged impurity is adjacent to the BLG, it creates a localized in-gap state. For positive impurities, a conduction band (CB) electron is localized at an energy $E_i^+$ within the gap, while for negative impurities a valence band (VB) hole is localized with energy $E_i^-$. We consider the isolated impurity energy $E_i^{\pm}$ to set the characteristic energy for mid-gap impurity states. Specifically, we consider a model in which in-gap localized states are distributed uniformly between the energy $E_i^{\pm}$ and the band edge, i.e., the density of states of the impurity band is taken to be constant. (In an actual device the distribution of energies for the in-gap states originates from the random positions of the ions in the IL, both in the sense that distance between the ``impurity'' ion and the BLG varies, and in the sense that larger-scale fluctuations of the IL create a random background potential). The total number $n_\textrm{imp}$ of localized in-gap states is a model parameter.

We envision that in our devices, in-gap states mainly originate from the interaction of charge carriers with ions in the ionic liquid (IL) electrolyte, in which ions with both charges are mobile. Because of this, at large perpendicular electric field ions of one polarity (whose sign depends on the sign of the interlayer energy difference $\Delta$) cover a large fraction of the interface with the BLG and determine the average potential, but ions with opposite charge are also present and create sharp potential fluctuations responsible for the formation of bound states (this is different from what happens at the interface between BLG and the Li-ion glass electrolyte, where only positive ions are mobile, so that at large electric field only ions of one polarity are present).  Inverting  the polarity of the interlayer potential $\Delta$ only causes the accumulation of ions of opposite polarity at the IL/BLG interface, and the behavior of the device is expected to be similar for the two polarities (because BLG is nearly perfectly electron-hole symmetric). The main difference between the two polarities is that positive and negative ions in the IL are molecules of different size, which can result in different density of in-gap states when applying interlayer potentials of equal magnitude but opposite polarity. We therefore describe the behavior of the chemical potential near the knee using a model in which all $n_\textrm{imp}$ localized states have the same sign. (Clearly, considering the simultaneous presence of bound states of both polarities would give more flexibility in modeling the evolution of the chemical potential, but it would require the introduction of more parameters).

\begin{figure}
    \centering
    \includegraphics[width=\columnwidth]{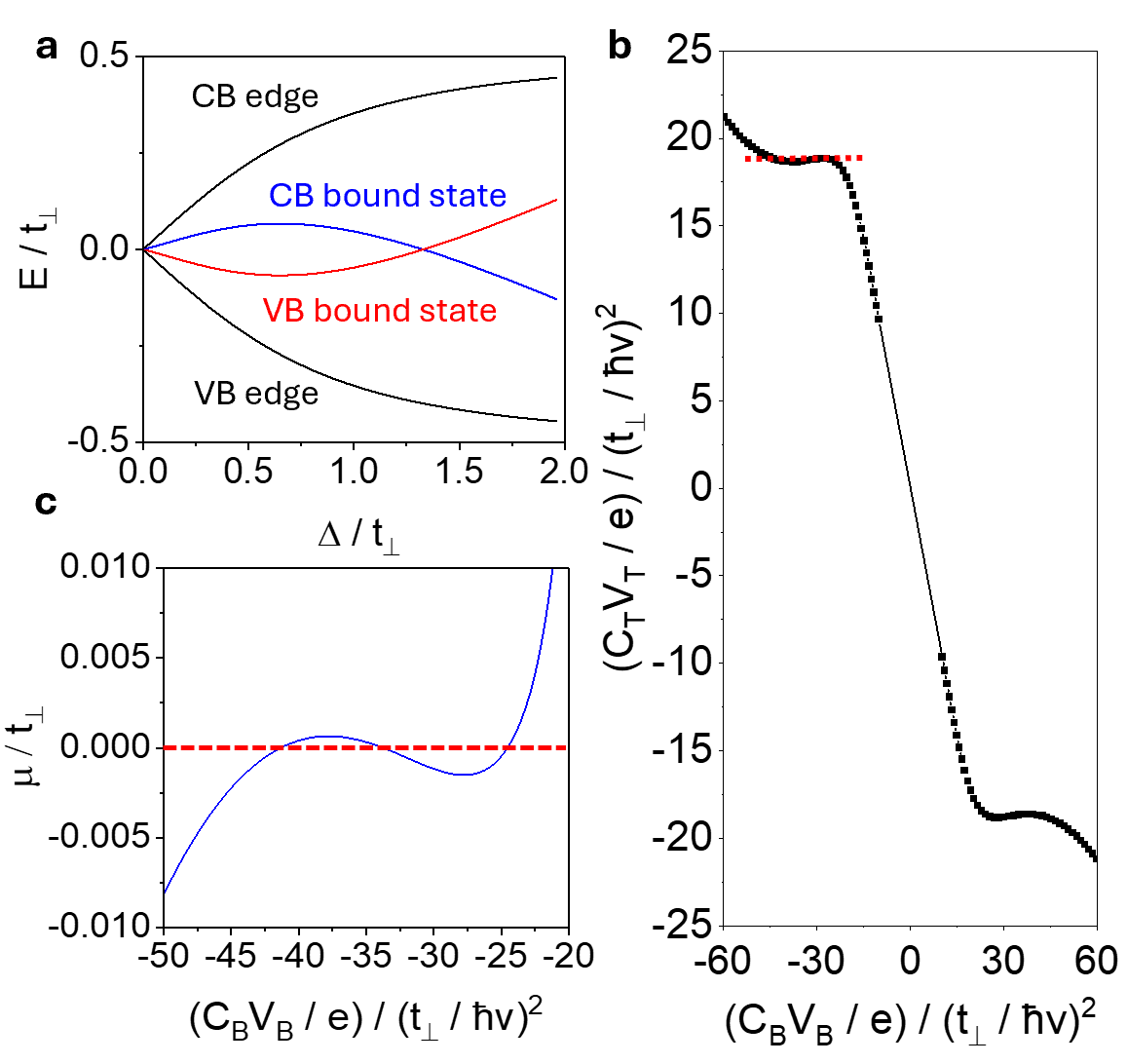}
      \caption{\justifying\small (a) Model calculation of the energy of an individual in-gap state as a function of the interlayer energy difference $\Delta$. Red and blue lines indicate the energy of electron and hole in-gap bound states created respectively by positive and negative charged ions in the ionic liquid electrolyte. The energy of the in-gap state exhibits a non-monotonic evolution with $\Delta$ and for a certain value $\Delta = \Delta_0$ it crosses through $E=0$ (the mid-gap level). Black lines indicate the band edges. (b) Solution of the equation $\mu(V_B,V_T)=0$  calculated with the model discussed in the main text. The resulting curve can be compared to the evolution of the position of the peak in $R_{xx\square}$ as a function of $\delta V_R^B$ and $\delta V_R^T$ (see Fig.\ref{fig:Figure3}a,b). The ``knees'' in the curve occur when  the edge of the impurity band passes through the mid-gap energy $E = 0$. (c) Non-monotonic dependence of the chemical potential as a function of $V_B$ for at a value of $V_T$  fixed in the vicinity of  the knee (see red dashed line in (b)). All calculations correspond to a temperature $k_B T = 0.025 t_\perp$ and use a dielectric constant $\varepsilon = 3$ and a net impurity concentration $n_\textrm{imp} = 100 (t_\perp/\hbar v)^2$, where $v$ is the velocity of Dirac fermions in monolayer graphene.  The symbols $C_B$ and $C_T$ denote the top gate and back gate capacitances, respectively. }
    \label{fig:Figure5tot}
\end{figure}

The ability of this simple model to capture the essence of the observed phenomenology arises because the energy $E_i^{\pm}$ of the in-gap state created by a Coulomb impurity depends non-monotonically on the interlayer potential $\Delta$ \cite{skinner_bound_2014, gorbar_electron_2024}, as depicted in Fig.~\ref{fig:Figure5tot}a. At small $\Delta \lesssim t_\perp$ -- as the band gap opens -- the bound state moves away from the mid-gap energy ($E = 0$). At $\Delta \gtrsim t_\perp$, however, the band gap saturates while the ionization energy of the bound state increases (due to the increasingly large density of states near the band edges), so that the in-gap state energy $E_i^{\pm}$ moves back towards $E = 0$. Eventually, at some sufficiently large value of $\Delta > \Delta_0$, the in-gap state energy crosses through $E = 0$. A precise calculation of the in-gap state energy is complicated by the large interband dielectric response  \cite{silvestrov_wigner_2017, joy_wigner_2022, joy_wigner_2023, gorbar_electron_2024}, which truncates the short-distance divergence of the Coulomb potential.  This effect is most prominent at small band gap, which is less relevant for understanding the experimental phenomenology  associated with the large band gap regime (note also that our experiments are at room temperature, so that the thermal energy is relatively large). For this reason, we are able to rely on a calculation of $E_i^{\pm}$ that uses an approximation of the screened potential at short distances (see Appendix \ref{sec:boundstate} for details).

The crossing of the in-gap state energy through the mid-gap level ($E=0$) implies an unusual dependence of the chemical potential on the two gate voltages. To see why this is so, let us imagine fixing the polarity of the interlayer potential $\Delta$ so that in-gap bound states originate from the valence band, just below $\Delta = \Delta_0$ ($\Delta_0$ is the value of $\Delta$ for which the bound state energy $E_i^-=0$). Under these conditions, at low temperature the state with chemical potential $\mu = 0$  corresponds to a vanishing density of charge $n = 0$ (i.e., to charge neutrality), because the in-gap energy is below $\mu$, and all valence band states are filled (irrespective of whether they are extended or bound to ionic charges). If we now increase  the interlayer potential $\Delta > \Delta_0$,  the in-gap states begin to pass through $E=0$, so that the condition $\mu = 0$ is associated with a net electron density that is negative, since some of the VB impurity states are empty. Varying the interlayer potential across $\Delta_0$ while keeping the BLG at charge neutrality, therefore, requires adjusting the chemical potential, moving it upward toward the conduction band. 

In general, the chemical potential for a given net charge density $n$ and a given temperature is determined by integrating the Fermi function across the energy-dependent density of states, which includes both itinerant and localized states (see Appendix \ref{sec:findmu} for details). The net electron density is related to the top-gate and back-gate voltages as
\begin{equation}
n = \frac{C_B V_B + C_T V_T}{e},
\label{eq:nVs}
\end{equation}
where $C_B$ and $C_T$ are the top-gate and back-gate capacitances, respectively. The value of the interlayer potential $\Delta$ is also fixed by the gate voltages according to 
\begin{equation}
\Delta = \frac{d}{2 \varepsilon \varepsilon_0}\left(1 - \frac{e \alpha}{\varepsilon \varepsilon_0}\right) (C_T V_T - C_B V_B)
\label{eq:nDelta}
\end{equation}
(as discussed above for Eq. (1), $\alpha$ quantifies electrostatic screening of the charges in the BLG).
In this way one can determine the chemical potential for a given set of gate voltages $V_B, V_T$.
Some results from our model calculation are shown in Fig.~\ref{fig:Figure5tot}b. The condition $\mu(V_B,V_T)=0$ gives the black line in the main figure, which exhibits a ``knee'' associated with the impurity band passing through $E=0$. The Fig.~\ref{fig:Figure5tot}c shows that at a fixed $V_T$ just above the ``knee", $\mu (V_B)$ is non-monotonic and crosses through zero multiple times. 

Our model gives a value of $\Delta_0 \approx 1.3 t_\perp \approx 0.45$ eV (Fig.~\ref{fig:Figure5tot}a). From this value one can estimate the voltage associated with the knee using Eq.~(\ref{eq:nDelta}) and setting $\Delta = \Delta_0$. 
Taking for both gate electrolytes a geometrical capacitance close to 50 $\mu$F/cm$^2$ as indicated earlier, we get a knee voltage $| V_B|, |V_T| \approx 0.5$ V. This is in good agreement  with the experimental value for $\delta V^R_B$(see Fig.~3). To also capture the experimental value of $\delta V^R_T$ with comparable precision we would need to consider that the capacitance of the two gate electrolytes are slightly different, and specifically that the capacitance to the top gate is --20-25 $\mu$F/cm$^2$-- somewhat smaller than the capacitance to the back gate. This is indeed what is  implied by the slope of the  peak in $R_{xx\square}$ at small gate voltage seen in Fig. 2a. Within our calculation, the horizontal extent of the flat region beyond the ``knee'' depends on the total number of impurity states, which for us is a model parameter. Once the charge density is large enough that a significant fraction of the impurity states are filled (e.g., $n \gtrsim n_\textrm{imp}$), the curve associated with $\mu = 0$ begins to disperse upward again. More in general, however, it is easy to understand that the detailed evolution of the $R_{xx\square}$ peak position as a function of the two gate voltages is determined by the energy dependence of the in-gap density of states, which is not constant as assumed here  (where  we have chosen the simplest option  to reproduce the key aspects of our experiments, to avoid introducing additional parameters).

In summary, our  model calculations confirm that the evolution of the bound state energy with increasing the interlayer potential has a large influence on the evolution of the chemical potential with gate voltage, in a way that accounts for the experimental phenomenology discussed in Section III. Even when limiting the parameters introduced to the barest minimum, our approach reproduces the correct magnitude of the gate voltages observed in the experiments (namely the position of the "knee"), which confirms the robustness of the phenomena discussed. As a last consideration, we point out an effect that may be important when analyzing the large interlayer potential regime in more detail, namely the contribution to the electrostatics given by  charges that occupy the in-gap states. After that these states cross the mid-gap level, the charges that occupy them are expected to increase the electrostatic potential difference between the layers in the BLG and not screen it (as is the case for charges accumulated in the band states in the absence of disorder). In the $\Delta \gg t_\perp$ regime, this effect may result in larger interlayer potential differences than naively expected, which may further amplify the effects of in-gap states on the chemical potential.

\section{Conclusions}
Our work shows that  double ionic gated Bernal-stacked bilayer graphene transistors enable the application of  a large perpendicular electric field, well above 1 V/nm, sufficient to  create an interlayer potential difference $\Delta$ that exceeds the interlayer hopping energy $t_\perp$. This regime could not be accessed experimentally until now and was therefore completely unexplored. Transport measurements  in this regime confirm that graphene bilayers exhibit a phenomenology different from that commonly observed in conventional devices with solid state dielectrics (in which dielectric breakdown of the gate insulators limits the maximum value of $\Delta$ to be well below $t_\perp$).

The measurements presented in this paper and their analysis have two main implications. First, they validate experimentally theoretical predictions about the evolution of the in-gap state energy at large interlayer potential, which could so far not be tested experimentally. These results are interesting not only because in the context of the physics of semiconductors such behavior of in-gap states is rather unique, but also also because excitons in gapped bilayer graphene --being bound states caused by Coulomb potentials-- are likely to share some of the unusual properties of the in-gap bound states discussed here \cite{park_tunable_2010, skinner_interlayer_2016}. Our work, therefore, calls for investigation of excitonic physics in bilayer graphene  at large  $\Delta$.  Second, the results presented here show that more systematic investigations of bilayer graphene in the $\Delta > t_{\perp}$ regime are possible.  In this first generation of double ionic gated devices, the effect of in-gap states was found to be dominant, and future work should target experiments in which ion-induced disorder is suppressed. This suppression can be accomplished by interposing atomically thin insulating layers (such as 2-3 layer hBN) between the electrolytes and the graphene bilayer \cite{gallagher_high-mobility_2015}. If sufficiently  thin, these layers allow the application of the large perpendicular electric fields needed to reach the  $\Delta > t_\perp$  regime, but would still space away the ions in the electrolyte, thereby suppressing substantially the resulting spatial potential variations. Including atomically thin hBN layers to separate the BLG from the electrolytes would therefore minimize  the density of in-gap states and give access to other physical phenomena originating from the drastic reshaping of the band structure of bilayer graphene at large interlayer potential.

\section*{Acknowledgments}
We gratefully acknowledge A. Ferreira for technical support, E. Linardy and C. Cao for contributions to the development of the technique of double ionic gating, V.Falko and A.Garcia-Ruiz for several discussions about the electronic properties of bilayer graphene at large perpendicular electric field. AFM acknowledges financial support from the Swiss National Science Foundation, Division II, under  grant 200021-227636. BS acknowledges support from the NSF under Grant No.~DMR-2045742.

\appendix

\section{Theoretical calculation of the bound state energy of an isolated impurity charge}
\label{sec:boundstate}

In order to estimate the energy of a mid-gap state created by a single isolated impurity charge, we follow the variational calculation described in Ref.~\cite{skinner_bound_2014}. Neglecting trigonal warping, one can approximate the energy of the CB as \cite{mccann_electronic_2013}
\begin{equation}
    E(k) = \left( \frac{t_\perp^2}{2} + \frac{\Delta^2}{4} + (\hbar v k)^2 - \sqrt{\frac{t_\perp^2}{4} + (\hbar v k)^2 (t_\perp^2 + \Delta^2)} \right)^{1/2}.
\end{equation}
This equation describes a ``moat'' or ``Mexican-hat'' band shape, such that the band edge occurs at a ring in momentum space with radius 
\begin{equation}
    k_0 = \frac{\Delta}{2 \hbar v} \sqrt{\frac{2 t_\perp^2 + \Delta^2}{t_\perp^2 + \Delta^2} }.
\end{equation}

The wave function for a low-energy state has momentum components that are mostly close in magnitude to $k_0$, or in other words the wave function in $k$-space is ring-shaped. This observation motivates a single-band variational wave function (in momentum space):
\begin{equation}
    \tilde{\psi}(k) = \frac{A}{1 + b^2 (k - k_0)^2}.
\end{equation}
The length scale $b$ represents the size of the wave function in real space and is the only variational parameter in the calculation. The constant $A$ is a ($b$-dependent) normalization constant.

The energy of (say) a CB bound state for a given value of $b$ is
\begin{align}
    E_b & = \int \frac{d^2 k}{(2 \pi)^2} \left| \tilde{\psi}(k) \right|^2 E(k)  \nonumber \\ 
    & - \int d^2 r V(r) \left| \psi(r) \right|^2,
    \label{eq:Evariational}
\end{align}
where $\psi(r)$ is the wave function in real space [the inverse Fourier transform of $\tilde{\psi}(k)$] and $V(r)$ is the Coulomb potential in real space. The first integral in Eq.~(\ref{eq:Evariational}) represents the kinetic energy and the second represents the potential energy of interaction with the impurity.

When the band gap is large compared to the ionization energy of the impurity, one can use an unscreened Coulomb potential $V(r) = V_0(r) = e^2/[4 \pi \varepsilon \varepsilon_0 r]$ to describe a coplanar impurity. However, when the band gap is narrow or the impurity energy is deep, the effect of interband dielectric screening becomes significant, and this screening modifies the form of $V(r)$. Specifically, such screening has the effect of truncating the $1/r$ divergence of the Coulomb potential at a distance shorter than the characteristic distance $r_0 = 8 e^2/(12 \pi \varepsilon_0 \Delta)$ \cite{silvestrov_wigner_2017, joy_wigner_2022, joy_wigner_2023}. While the exact form of the screened interaction is complicated and does not admit a simple expression for $V(r)$, we can approximate the screening effect by using the substitution
\begin{equation}
    V(r) = \left[ \frac{1}{V_0(r)} + \frac{1}{V_0(r_0)} \right]^{-1},
\end{equation}
so that in effect the short-distance divergence of the Coulomb potential is truncated at $V(r_0) = (3/8) \Delta/\varepsilon$. 

For a given value of the parameter $\Delta$, we estimate the bound state energy as that which minimizes the total energy with respect to the variational parameter $b$:
\begin{equation}
    E_i \simeq {\min_{b}} \ E_b.
\end{equation}
The result of this calculation is shown in Fig.~\ref{fig:Figure5tot}a. Notice that at $\Delta = t_\perp$ the value of $r_0 \sim e^2/(4\pi\varepsilon_0 t_\perp) \sim 4$\,nm. Since this distance is much longer than the spacing of the ions from the plane of the BLG, the impurity state energy is effectively insensitive to the distance of the ion from the BLG and we can effectively set it to zero in our calculations.

\section{Theoretical calculation of the chemical potential as a function of the top-gate and back-gate voltages}
\label{sec:findmu}

The gates voltages are related to the net electron density $n$ and the interlayer potential $\Delta$ by Eqs.~(\ref{eq:nVs}) and (\ref{eq:nDelta}).
For a given value of $n$ and $\Delta$, we can find the chemical potential by searching numerically for the solution to the equation
\begin{align}
n = & \int_{-\infty}^{\infty} f(E; \mu, T) g_\text{CB}(E) dE \nonumber \\
& - \int_{-\infty}^{\infty} [1-f(E; \mu, T) ]g_\text{VB}(E) dE,
\label{eq:nintegral}
\end{align}
where $f(E; \mu, T) = [1 + \exp((E-\mu)/k_B T)]^{-1}$ is the Fermi function. The first integral on the right-hand side of Eq.~(\ref{eq:nintegral}) represents the number of electrons in the conduction band and the second integral represents the number of holes in the valence band. The quantity $g_\text{CB}(E)$ represents the total density of states for conduction band electrons, including both the mid-gap impurity states and the bare conduction band states; $g_\text{VB}(E)$ is the analogous quantity for the valence band.

For the black curve shown in Fig.~\ref{fig:Figure5tot}b, we have set the impurity band DOS to be a constant $n_\textrm{imp}/|E_i - E_\textrm{edge}|$ for energies between $E_i$ and the band edge and zero elsewhere, where $E_\textrm{edge}$ is the energy of the CB edge for $\Delta < 0$ (positive impurities dominate) and the VB edge for $\Delta > 0$ (negative impurities dominate).

%%%%%%%%%%%%%%%%%%%%%%%%%%%%%%%%%%%%%%%%
\newpage
\input{References.bbl}
\end{document}